\documentclass[utf8]{frontiersFPHY} % for Physics and Applied Mathematics and Statistics articles

\setcitestyle{square} % for Physics and Applied Mathematics and Statistics articles
\usepackage{url,hyperref,microtype,subcaption}
\usepackage[onehalfspacing]{setspace}
\usepackage{amsmath, amssymb, bbm}
\usepackage{booktabs,multirow}
\usepackage{graphicx}
\usepackage{makecell}

\makeatletter
\def\@evenfoot{\vbox to 12.5\p@{\rule{\textwidth}{0.5\p@}\vss
        \hbox to \textwidth{\helvetica\small {}\hfill\helveticabold\color{black}\thepage}%
        }}%

\def\@oddfoot{\vbox to 12.5\p@{\rule{\textwidth}{0.5\p@}\vss
        \hbox to \textwidth{\helveticabold\small {}\hfill \thepage}%
        }}

\def\@maketitle{%
        \let\footnote\thanks
        \clearemptydoublepage
        \checkoddpage\ifcpoddpage\setlength{\aboveskipchk}{-7pc}\else\setlength{\aboveskipchk}{-3pc}\fi%

        \vspace*{\aboveskipchk}%
        \vspace{\dropfromtop}%
        \hbox to \textwidth{}

        \helvetica
        \hbox to \textwidth{%
        \parbox[t]{36.5pc}{%
          \vspace*{1sp}
          {\helveticabold\fontsize{20}{21}\color{black}\selectfont\raggedright \@title \par}%
          \vspace{4.5\p@}
          {\helveticabold\fontsize{12}{15}\selectfont\raggedright \@author \par}%
              \vspace{4\p@}
          {\helvetica\fontsize{12}{12}\selectfont\raggedright\slshape\@address \Address \\ \par}%
            \vspace{6\p@}
          {\helvetica\fontsize{12}{12}\selectfont\raggedright\@extraAuth \par}%
          \vspace{8\p@}
          }%
        }
        \vspace{14.5\p@}%
        }

\makeatother

\renewcommand{\vec}[1]{\textbf{#1}}
\newcommand{\tensor}[1]{\mathbbm{#1}}
\newcommand{\diff}[1]{\mathrm{d}#1}
\newcommand{\sym}{\operatorname{sym}}
\newcommand{\tr}{\operatorname{tr}}
\newcommand{\abs}[1]{\left|#1\right|}

\def\keyFont{\fontsize{8}{11}\helveticabold }
\def\firstAuthorLast{Lautenbach {et~al.}} %use et al only if is more than 1 author
\def\Authors{S. Lautenbach\,$^{1}$ and R. Grauer\,$^{1,*}$}
\def\Address{$^{1}$Institut f\"ur Theoretische Physik I, Fakult\"at f\"ur Physik und Astronomie, Ruhr-Universit\"at Bochum, Germany}
\def\corrAuthor{Rainer Grauer \\ Institut f\"ur Theoretische Physik I, Fakult\"at f\"ur Physik und Astronomie, Ruhr-Universit\"at Bochum, Universitätsstr. 150, D-44801 Bochum, Germany}
\def\corrEmail{grauer@tp1.rub.de}

\begin{document}
\onecolumn
\firstpage{1}

\title[Multiphysics simulations of collisionless plasmas]{Multiphysics simulations of collisionless plasmas}

\author[\firstAuthorLast ]{\Authors} %This field will be automatically populated
\address{} %This field will be automatically populated
\correspondance{} %This field will be automatically populated

\extraAuth{}% If there are more than 1 corresponding author, comment this line and uncomment the next one.
%\extraAuth{corresponding Author2 \\ Laboratory X2, Institute X2, Department X2, Organization X2, Street X2, City X2 , State XX2 (only USA, Canada and Australia), Zip Code2, X2 Country X2, email2@uni2.edu}

\maketitle

\begin{abstract}

%%% Leave the Abstract empty if your article does not require one, please see the Summary Table for full details.
% \section{}
% For full guidelines regarding your manuscript please refer to
% \href{http://www.frontiersin.org/about/AuthorGuidelines}{Author Guidelines}.
%
% As a primary goal, the abstract should render the general significance and
% conceptual advance of the work clearly accessible to a broad
% readership. References should not be cited in the abstract. Leave the Abstract
% empty if your article does not require one, please see
% \href{http://www.frontiersin.org/about/AuthorGuidelines#SummaryTable}{Summary
%   Table} for details according to article type.

  Collisionless plasmas, mostly present in astrophysical and space environments,
  often require a kinetic treatment as given by the Vlasov equation.
  Unfortunately, the six-dimensional Vlasov equation can only be solved on very
  small parts of the considered spatial domain. However, in some cases, e.g.~magnetic
  reconnection, it is sufficient to solve the Vlasov equation in a
  localized domain and solve the remaining domain by appropriate fluid models.
  In this paper, we describe a hierarchical treatment of collisionless plasmas
  in the following way. On the finest level of description, the Vlasov equation
  is solved both for ions and electrons. The next courser description treats
  electrons with a 10-moment fluid model incorporating a simplified treatment of
  Landau damping. At the boundary between the electron kinetic and fluid region,
  the central question is how the fluid moments influence the electron
  distribution function. On the next coarser level of description the ions are
  treated by an 10-moment fluid model as well. It may turn out that in some
  spatial regions far away from the reconnection zone the temperature tensor in
  the 10-moment description is nearly isotopic. In this case it is even possible
  to switch to a 5-moment description. This change can be done separately for
  ions and electrons. To test this multiphysics approach, we apply this full
  physics-adaptive simulations to the Geospace Environmental Modeling (GEM)
  challenge of magnetic reconnection.

\tiny
\keyFont{ \section{Keywords:} multiphysics coupling, kinetic plasmas, fluid descriptions, numerical simulations, reconnection}
% All article types: you may provide up to 8 keywords; at least 5 are mandatory.
\end{abstract}

\section{Introduction}
One of the most important challenges in astrophysical, space and fusion plasmas
is the treatment of different spatial and temporal scales and the correct
physical description on each of these different scales.

In order to give a rough estimate for different plasma systems, let us first
consider the warm ionized phase (diffuse ionized hydrogen) in the interstellar
medium. Here, the smallest relevant kinetic scales are in the order of magnitude
of kilometres, while the global scale of the system is about $10^{13}$\,km. In
the heliosphere the scales are altogether smaller (kinetic scales about $2$\,km,
system scale about $10^8$\,km), but the ratio of global to kinetic scales is
still astronomical in the truest sense. The situation is similar in fusion
plasmas: the electron skin depth is about $5\cdot 10^{-4}$\,m and the vessel measures
about $10$\,meters. In all these cases, it is not possible to carry out
simulations which represent all scales with the finest level (kinetic equations)
of the physical description.
Most of these plasmas can be considered as collisionless, since collision times
are orders of magnitude larger than time scales relevant for the dynamical
evolution of the plasma. Such plasmas can be modelled with the kinetic Vlasov
equation.  Nevertheless, kinetic models are inherently computationally
expensive, so that large--scale simulations of typical phenomena, as for example
magnetic reconnection or collisionless shocks, are hardly feasible and only
possible in localized regions of interest.
As an alternative, much cheaper fluid models can be considered, but they lack
the expressiveness and some physics of full kinetic models, even though some of
the effects may be included. Simple treatments and modelling of Landau damping
in the same context were proposed and analyzed in
\cite{wang-et-al:2015,ng-hakim-etal:2017,allmann-rahn-trost-grauer:2018}. These studies were based
on the closure introduced by \citet{hammett-perkins:1990} and successive work in
this direction \citep{hammett-dorland-perkins:1992,passot-sulem:2003}. An
extension providing heat fluxes in the parallel and perpendicular directions
(with respect to the magnetic field) was presented in
\cite{sharma-hammett-etal:2003}. An excellent overview is given in
\citet{chust-belmont:2006}.

Fortunately, many relevant problems like magnetic reconnection or collisionless
shocks exhibit a rather clear separation of scales and regimes such that an
adaptive approach is promising and might combine the best of the two worlds:
cheap models where they are sufficient and detailed models where they are
necessary and interesting.
The idea of coupling different physical models is not new and has been applied
in different physical contexts. Schulze et al. \cite{Schulze2003} couple kinetic
Monte-Carlo and continuum models in the context of epitaxial growth.
Considerable efforts have been made to couple kinetic Boltzmann descriptions
with fluid models (see e.g.
\cite{deg2010,del2003,gou2013,tiwari-klar:1998,Tal1997}).
In the context of plasma physics Sugiyama and Kusano \cite{Sug2007}, Markidis et
al. \cite{Mar2014} and Daldorf et al. \cite{daldorff-et-al:2014} show ways to
combine PIC and MHD fluid models, and Kolobov and Arslanbekov \cite{Kol2012}
describe the transition from neutral gas models to models of weakly ionized
plasmas.

We take a slightly different route in solving the Vlasov equation on the finest
relevant scales and then adaptively use less and less detailed fluid models
outside the kinetic region. In this way we have some control where to use which
kind of physical model at the expense of dealing with a substantially more
complicated computational infrastructure.

Our group has developed and is continuously developing and improving methods and
codes that are capable of combining kinetic and fluid models during runtime
\cite{rieke-et-al:2015}, making it possible to consider problems of the type
mentioned above at much lower expenses than before.

\begin{figure}[h!]
  \centerline{\includegraphics[width=0.9\textwidth]{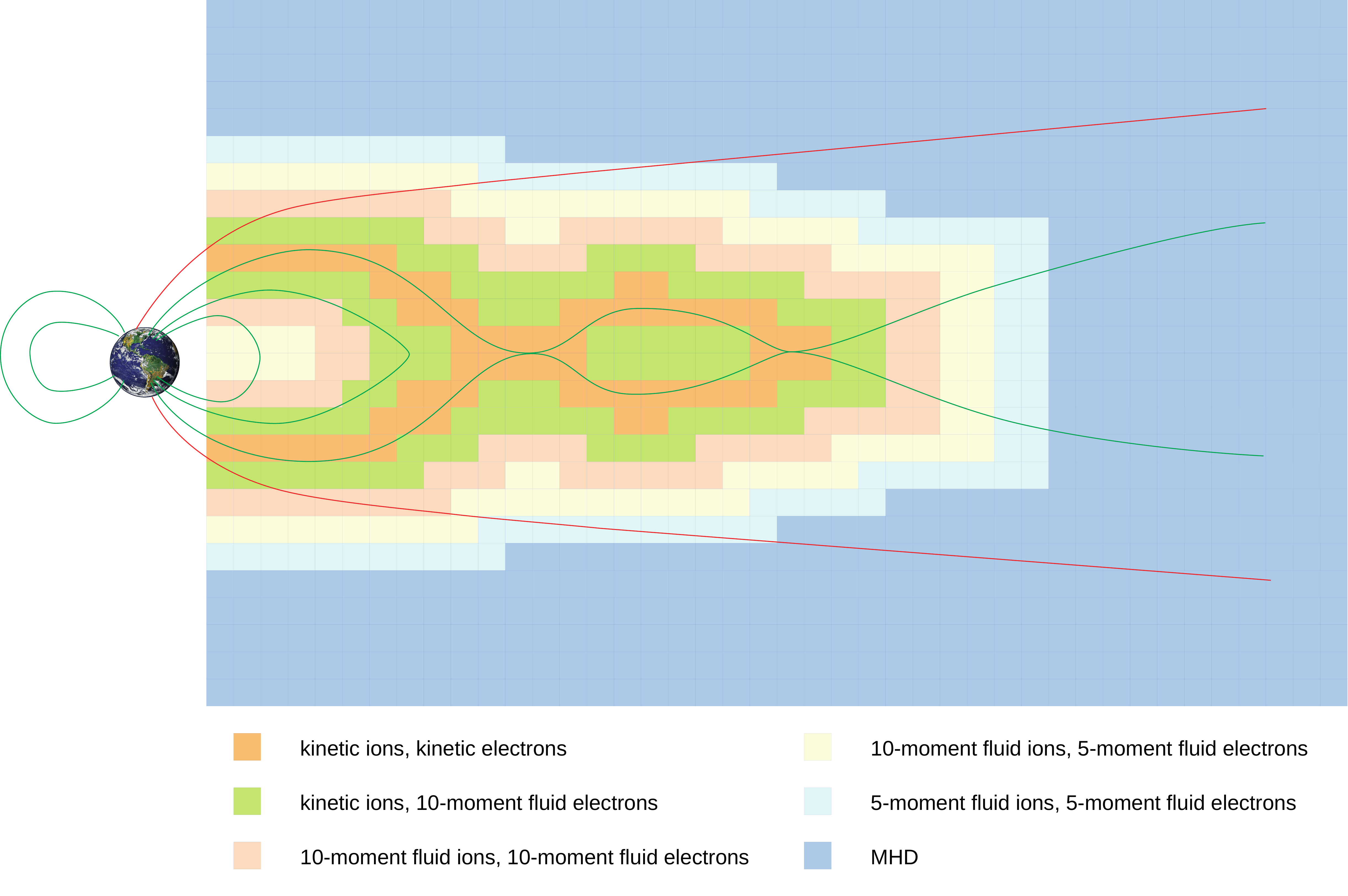}}
  \caption{Oversimplified sketch of a multiphysics approach for tail reconnection}
  \label{fig:sketch}
\end{figure}

A sketch of this hierarchy is depicted in figure~\ref{fig:sketch}. In the inner
zone, both ions and electrons are treated kinetically and solved with the
Vlasov equation. Adjacent to this zone, ions are still modelled with the Vlasov
equation but electrons are described with a 10-moment fluid model. On the next
coarser level of description, the ions are also described by a 10-moment fluid
model. To ease the transition from the kinetic to the 10-moment fluid
description we apply the Landau closure developed in
\cite{wang-et-al:2015} in the fluid description.
It may turn out that in some spatial regions outside the reconnection zone the
temperature tensor in the 10-moment description is nearly isotopic. In this case
it is even possible to switch to a 5-moment description. This change can be done
separately for ions and electron.
In future studies we will also try to include the coupling of the 5-moment model
to magnetohydrodynamic (MHD) models (with generalised Ohms law) which would
represent the last step in this hierarchy.

With this multiphysics strategy, these codes can be applied to problem sizes
that are otherwise impossible to reach with kinetic simulations and the
understanding of the impact of small scale phenomena on the dynamics on global
scales is in reach.

The outline of the paper is the following: first we briefly describe all the
plasma models and the necessary numerical schemes (Vlasov equation, 10- and
5-moment fluid equations, Maxwell's equations, the coupling procedure, the
Landau fluid closure). We will then study the Geospace Environmental Modeling
(GEM) reconnection setup \cite{birn2001} and perform comparisons to pure kinetic
and pure fluid simulations.

\section{Plasma Models}

The plasma models that we have to consider are: i) the Vlasov equation,
ii) Maxwell's equations and iii) the 10- and 5-moment fluid equations. We will briefly
summarise these sets of equations.

\subsection{Vlasov equation}
Collisionless plasmas on the finest level of description are governed by the Vlasov equation
\begin{equation}
  \label{eq:vlasov-eq}
  \partial_t f_s(\vec{x},\vec{v},t) + \vec{v}\cdot \nabla_{\vec{x}}f_s(\vec{x},\vec{v},t)
  + \frac{q_s}{m_s}\big(\vec{E} + \vec{v} \times \vec{B} \big)\cdot \nabla_{\vec{v}}f_s(\vec{x},\vec{v},t) = 0\;,
\end{equation}
where $f_s(\vec{x},\vec{v},t)$ denotes the phase-space density, $q_s$ and $m_s$ the
particle charge and mass for species $s \in \{e,i\}$ (electrons and ions).
The electric and magnetic fields $\vec E$ and $\vec B$ are given by Maxwell's equations:
\begin{subequations}\label{eq:maxwell-eq}
  \begin{align}
    \nabla \cdot \vec E &= \frac{\rho}{\varepsilon_0} \\
    \nabla \cdot \vec B &= 0 \\
    \partial_t \vec B &= - \nabla \times \vec E \label{eq:faradays-law} \\
    \partial_t \vec E &= c^2\left(\nabla\times\vec B - \mu_0 \vec j\right) \label{eq:amperes-law}
  \end{align}
\end{subequations}
with speed of light $c$ and electric constant $\varepsilon_0$. Maxwell's
equations depend on charge and current densities $\rho$ and $\vec j$, which are
obtained from the phase-space densities $f_s(\vec{x},\vec{v},t)$:
\begin{subequations}
  \begin{align}
    \rho &:= \sum_s q_s \int\! f_s(\vec{x},\vec{v},t)\,\diff^3 v, \label{eq:charge-density} \\
    \vec j &:= \sum_s q_s \int\! \vec v f_s(\vec{x},\vec{v},t) \,\diff^3 v\;. \label{eq:current-density}
  \end{align}
\end{subequations}
Vlasov equation \eqref{eq:vlasov-eq} and Maxwell's equations \eqref{eq:maxwell-eq}
form a closed set of equations and constitute the most fundamental description
of a collisionless plasma.

\subsection{Two-species fluid equations}
Fluid descriptions can be obtained from the Vlasov equation \eqref{eq:vlasov-eq} by taking
moments of the phase-space density $f_s$,
\begin{equation}\label{eq:def-nth-moment}
  \mu_{n,s} := \int\! \vec{v}^n f_s(\vec{x},\vec{v},t) \,\diff^3 v \, .
\end{equation}
Here, $\vec{v}^n$ denotes the n-fold tensor product of $\vec v$ with itself, $\vec{v}^0 := 1$.
Typically, only the first few moments are considered since a Gaussian
distribution $f_s(\vec{x},\vec{v},t)$ is exactly represented by the moments $\mu_{0,s}$,
$\mu_{1,s}$, $\mu_{2,s}$  (and all other moments equaling zero).

We will subsequently the describe the 10- and 5-moment equations. Consider the lowest
moments up to $\mu_{3,s}$ :
\begin{subequations}
  \begin{flalign}
    \label{eq:mass-density}
    &\text{particle density:} & n_s &:= \mu_{0,s} = \int\! f_s(\vec{x},\vec{v},t) \,\diff^3 v &\\
    \label{eq:momentum-density}
    &\text{bulk velocity:} & \hat{\vec u}_s &:= \frac{\mu_{1,s}}{\mu_{0,s}}  = \frac{1}{n_s}\int\! \vec v f_s(\vec{x},\vec{v},t) \,\diff^3 v &\\
    \label{eq:energy-tensor}
    &\text{energy density tensor:} & \tensor{E}_s &:= m_s \mu_{2,s} = m_s\int\! \vec v^2 f_s(\vec{x},\vec{v},t) \,\diff^3 v &\\
    \label{eq:heat-flux}
    &\text{heat flux tensor:} & \tensor{Q}_s &:= m_s \mu_{3,s} = m_s\int\! \vec v^3 f_s(\vec{x},\vec{v},t) \,\diff^3 v &
  \end{flalign}
\end{subequations}
$n_s$, $\hat{\vec u}_s$, $\tensor E_s$ are evolved by the following equations,
obtained from the Vlasov equation \eqref{eq:vlasov-eq}:
\begin{subequations}\label{eq:10mom-eq}
  \begin{align}
    \partial_t n_s =& - \nabla \cdot (n_s\hat{\vec u}_s) \label{eq:continuity-3d} \\
    \partial_t (m_sn_s\hat{\vec u}_s) =& - \nabla\cdot\tensor{E}_s + q_s \big( n_s\vec{E} + n_s\hat{\vec u}_s \times \vec B \big) \label{eq:momentum-3d} \\
    \partial_t \tensor{E}_s =& -\nabla\cdot\tensor{Q}_s + 2 q_s \sym\!\left(n_s\hat{\vec u}_s \vec E + \frac 1 {m_s}\tensor{E}_s\times \vec B\right) \label{eq:energy-3d}
  \end{align}
\end{subequations}
Naturally, these equations are not closed. Designing appropriate fluid closures
have a long history. An excellent overview is given in
\citet{chust-belmont:2006}. In order to mimic kinetic Landau damping effects,
several closures have been developed (see
\citep{hammett-dorland-perkins:1992,passot-sulem:2003}), all based on the early
\citet{hammett-perkins:1990} model.

Wang et al. \cite{wang-et-al:2015} suggested a heat flux closure which
approximates a spectrum of wave numbers by one single wave number $k_0$.
Following this idea, Allmann-Rahn et al.  \cite{allmann-rahn-trost-grauer:2018}
developed an improved model that is able to correctly describe the kinetic
scaling of average reconnection rate $(\lambda/d_{i})^{-0.73}$ as a function of
the distance between the islands' $O$-points $\lambda$ and where $d_i$ denotes the ion
skin depth (see figure 9 in \cite{allmann-rahn-trost-grauer:2018}).

In the simulations used in this paper the original $k_0$-closure from
\citet{wang-et-al:2015} is used. It approximates the divergence of the heat flux
with an expression that forces an anisotropic pressure tensor to a more
isotropic one. More precisely, the expression reads
\begin{equation}\label{eq:closure-ten-moment}
  \nabla \cdot \tensor{Q}_s = v_{\mathrm{th},s}|k_0|(\tensor{P}_s - p_s\tensor 1)\;,
\end{equation}
with the pressure tensor $\tensor{P}_s = \tensor{E}_s - m_s n_s \hat{\vec
u}_s\hat{\vec u}_s$, the scalar pressure $p_s = \frac 1 3 \tr \tensor{P}_s$ and
thermal velocity $v_{\mathrm{th}}$.  The parameter $k_0$ is choosen on the order
of the inverse Debye length. Together with \eqref{eq:10mom-eq}, this constitutes a
closed set of ten fluid equations.

In future simulations it planed to switch to the improved model introduced in
\citet{allmann-rahn-trost-grauer:2018}.

As an alternative to the 10-moment description, an even simpler 5-moment
description can be introduced where the energy density tensor and heatflux
tensor are replaced by the scalar energy density $\mathcal{E}_s = \frac 1 2 \tr
\tensor{E}_s$ and vector heat flux $\vec Q_s$. The scalar energy density evolves
in time according to:
\begin{equation} \label{eq:energy-1d}
  \partial_t\mathcal{E}_s = -\nabla\cdot\vec{Q}_s - \nabla\cdot\left(\frac{5} 2 p_s\vec{u}_s - \frac 1 2 m_sn_s(\vec{u}_s\cdot\vec{u}_s)\vec{u}_s \right) + q_sn_s\vec{u}_s\cdot \vec E
\end{equation}
Together with the assumption of adiabaticity, $\nabla\cdot\vec{Q}_s \equiv 0$\;,
(\ref{eq:continuity-3d},~\ref{eq:momentum-3d},~\ref{eq:energy-1d}) form the set of
five moment equations.

Completely analogous to the case of the Vlasov equation, the source terms $\rho$
and $\mathbf{j}$ in Maxwell's equations~\eqref{eq:maxwell-eq} are formed from
the particle densities $n_s$ (see equation \eqref{eq:mass-density}). Actually,
as will be described in section \ref{sec:maxwell}, only the current density
$\mathbf{j}$ is needed to propagate Maxwell's equations.

\section{Numerical Methods}
\subsection{Vlasov equation}

In order to circumvent the complexity that could arise from the high
dimensionality of the phase space, the Vlasov equation is split into five
one-dimensional problems using Strang splitting \cite{strang:1968}. These
one-dimensional advection problems are solved with a third order semi-Lagrangian
flux-conservative scheme introduced by \citet{filbet2001}. In order to minimize
the error due to the Strang splitting when calculating the backward
characteristics needed in the semi-Lagrangian method, the cascade interpolation
\cite{leslie-purser:1995} is combined with the Boris step \cite{boris:1970} to
form the backsubstitution method introduced in~\citet{schmitz-grauer:2006a}.
Details of this procedure can be found in \cite{schmitz-grauer:2006b}.

The code is fully parallelized using the message passing interface (MPI)
\cite{MPI:1994} where the Vlasov part is solved in parallel on distributed
graphics cards using CUDA programming tools \cite{programming-guide}.

\subsection{Two-species fluid equations}

Both the 10-moment and the 5-moment two-species fluid models are
all discretized with the same numerical methods.

For the discretization in space, we use the CWENO scheme introduced by
\citet{kurganov-levy:2000}, an easy to implement third order finite-volume
scheme which is a perfect compromise between sharp shock resolution and
high-order approximation in spatially smooth regions. A third order
strong-stability-preserving Runge-Kutta scheme \cite{shu-osher:1988} is employed
for the time integration.

\subsection{Maxwell's equation} \label{sec:maxwell}

The electromagnetic fields are positioned on a staggered Yee grid
\cite{yee:1966} in order to maintain the divergence free condition for the
magnetic field: $\nabla \cdot \mathbf{B} = 0$. Equations (\ref{eq:faradays-law},
\ref{eq:amperes-law}) are evolved through the FDTD method presented in Taflove
and Brodwin~\cite{taflove1995}. Here, only the current density $\mathbf{j}$
enters as a source term. Since the speed of light exceeds all other speeds found
in the plasma by far, subcycling is used in order to resolve lightwaves while
keeping the global timestep as large as possible. In addition, the speed of
light is artificially reduced to $20$ times the Alfv\'en speed.

\subsection{Adaptive Coupling}

The coupling strategy is the most important and at the same time the most
critical part of the multiphysics simulations. The coupling strategy involves
two separate problems: first, providing the correct boundary conditions at
interfaces between different physical models and second, designing criteria to
decide which model can be used in which part of the computational domain in an
adaptive way.

We start with discussing the strategy for obtaining boundary conditions at the
interfaces. Providing boundary conditions for the fluid part at the
kinetic/fluid interface is rather straightforward: the fluid boundary conditions
are obtained by taking the necessary moments of the phase-space densities $f_s$
at the interface. Providing boundary conditions for the phase-space densities
$f_s$ from the fluid description is far less trivial and described in detail in
\citet{rieke-et-al:2015}. In short the procedure can be summarised as follows:
we first extrapolate the phase-space density $f_s$ to the boundary region. Next,
we adjust the extrapolated phase-space density $f_s$ such that the moments equal
the moments from the 10-moment fluid description. In this way, we only
manipulate the phase-space density $f_s$ rather ``smoothly'' with minimal changes
and do not force $f_s$ to a Gaussian shape.
The coupling between the 10-moment and 5-moment fluid regions is done in a very
natural way. The boundary conditions for the 5-moment description is simply
obtained by calculating the energy scalar from the trace of the energy density
tensor. In the other direction, the energy tensor has to be constructed from the
energy scalar by assuming a diagonal shape at the 10-/5-moment interface.

The criteria to decide which of the available models shall be used in which
subregion of the domain is a highly non-trivial issue. Presently, our strategy
is still in a phase of proof of concept and further work has to be invested. For
the case of magnetic reconnection, we implemented heuristic criteria based on
the current density $j_z$ since it is a good indicator for regions of high
reconnection. In order to allow for a finer detachment of electrons and ions, we
actually use the velocity $u_{z,s} = n_s \hat u_{z,s}$ for electrons and ions
separately. This is reasonable as the current density is given by $j_z = \sum_s
q_s u_{z,s}$ (compare \eqref{eq:current-density}).
It should be stated clearly that a criterion based on the current density is
rather heuristic and far from universally applicable approach and further work
has to be invested on robust criteria.
% A more general but not sufficient criterion might be the occurrence of high
% temperature gradients.

In addition, once a criterion is considered satisfactory for the context,
thresholds have to be defined that mirror a good trade-off between the need to
use a higher-information model for a correct representation and the opportunity
to save computational resources with a lower-information model. Up to know we
can only state that this is based on educated guesses.

\renewcommand\theadalign{tl}
\begin{table}[h!]
  \caption{\label{tab_scaling} Scaling behavior of {\it muphy} on JURECA across
    different problem sizes (absolut number of grid cells) and number of
    GPUs. Given are the times needed for 2000 steps (in hours) and the
    relative speedup normalized to $2^{30}$ cells on one GPU. \label{tab:performance}}
\begin{center}
\setlength{\tabcolsep}{10pt}
\begin{small}
  \renewcommand{\arraystretch}{2.5}
  \begin{tabular}{c|cccccccc}
   \thead{\#\,GPUs\ $\rightarrow$\\ \#\,Cells\ $\downarrow$ }
   & $2^0$ & $2^1$ & $2^2$ & $2^3$ & $2^4$ & $2^5$ & $2^6$ & $2^7$ \\
   \hline
   $2^{26}$
   & \makecell{ 1.27\,h\\\textit{(14.7)}} & \makecell{ 0.59\,h\\\textit{(31.4)}} &
   &                                      &                                      &
   &                                      &                                      \\
   $2^{28}$
   & \makecell{ 4.60\,h\\\textit{(4.06)}} & \makecell{ 2.40\,h\\\textit{(7.80)}} & \makecell{ 1.23\,h\\\textit{(15.2)}}
   & \makecell{ 0.62\,h\\\textit{(29.9)}} &                                      &
   &                                      &                                      \\
   $2^{30}$
   & \makecell{ 18.7\,h\\\textit{(1.00)}} & \makecell{ 9.71\,h\\\textit{(1.92)}} & \makecell{ 4.92\,h\\\textit{(3.79)}}
   & \makecell{ 2.76\,h\\\textit{(6.77)}} & \makecell{ 1.25\,h\\\textit{(15.0)}} & \makecell{ 0.70\,h\\\textit{(26.63)}}
   &                                      &                                     \\
   $2^{32}$
   &                                      &                                      & \makecell{ 20.6\,h\\\textit{(0.91)}}
   & \makecell{ 10.0\,h\\\textit{(1.86)}} & \makecell{ 4.97\,h\\\textit{(3.75)}} & \makecell{ 2.62\,h\\\textit{(7.12)}}
   & \makecell{ 1.32\,h\\\textit{(14.1)}} & \makecell{ 0.69\,h\\\textit{(27.2)}} \\
   $2^{34}$
   &                                      &                                      &
   &                                      & \makecell{ 20.8\,h\\\textit{(0.90)}} & \makecell{ 10.8\,h\\\textit{(1.74)}}
   & \makecell{ 5.24\,h\\\textit{(3.57)}} & \makecell{ 2.83\,h\\\textit{(6.61)}}
  \end{tabular}
\end{small}
\end{center}
\end{table}
\begin{figure}[h!]
\begin{center}
 \includegraphics[width=\textwidth]{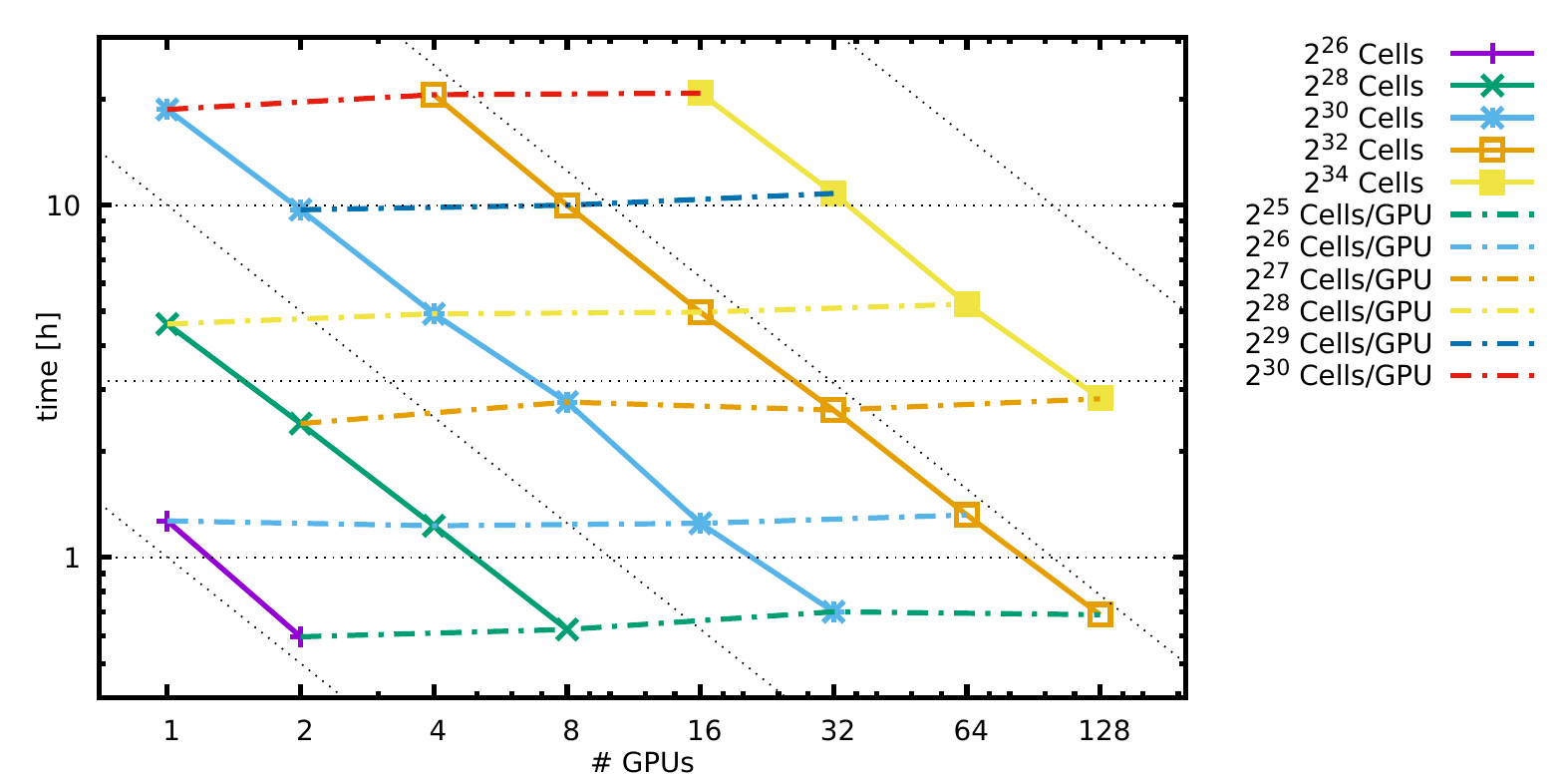}
\end{center}
\caption{Scaling behavior of {\it muphy} on JURECA. Same data as in table \ref{tab:performance}.
        \label{fig:performance} }
\end{figure}
\subsection{Code performance}

The described numerical codes, the adaptive coupling procedures and the
parallelisation framework based on space-filling curves
\cite{dreher-grauer:2005} is build in our framework called \textit{muphy}. This
framework has been developed over the last 10 years. It is
written in C++/CUDA, runs partly on GPUs and partly on CPUs and employs MPI for
parallelization.

Scaling runs have been performed on the JURECA supercomputer at the FZ J\"ulich,
Germany \cite{jureca}, on a fully kinetic Whistler-wave setting
\cite{daldorff-et-al:2014}. Scaling results are excellent as shown in table
\ref{tab:performance} and figure \ref{fig:performance}. Note that the number of
GPUs was only restricted by the actual configuration of JURECA.

\section{Results}
% This section may be divided by subheadings. Footnotes should not be used and
% have to be transferred into the main text.

We apply the described models and the multiphysics coupling strategy to the
Geospace Environmental Modeling (GEM) challenge \cite{birn2001}. The domain size
is chosen as $4\pi\, d_i$ in $x$- and as $2\pi\, d_i$ in
$y$-direction, where $d_i$ denotes ion skin depth. The symmetry properties
of the GEM problem make it sufficient to calculate only one quarter of the
spatial domain. We use a uniform cell-width of d$x =\; $d$y = \frac{\pi}{128}\,
d_i$ in 2d physical space and a uniform resolution of $32$ cells in each
direction in 3d velocity space. To reduce the computational costs, we apply the
common values for the reduced mass ratio $\frac{m_i}{m_e} = 25$ and reduced
speed of light of $20$ times the Alfv\'en speed. The numerical setup is depicted
in table \ref{tab:gem-setup}.
\begin{table}[h!]
  \caption{\label{tab:gem-setup} %
    Numerical setup of GEM}
  \renewcommand{\arraystretch}{1.2}
  \begin{tabular}{l l}
    \toprule %\midrule
    dimensions of physical domain & $\{x,y\} \in \{[-2\pi..2\pi],[-\pi..\pi]\}\,d_i$ \tabularnewline
    cell-width of physical space & d$x = $d$y = \frac{\pi}{128}\,d_i$ \tabularnewline
    size of subregions in physical space & $\tilde N_x = \tilde N_y = 32$\,cells \tabularnewline
    resolution of velocity space (kinetic region) & $N_{v_x} = N_{v_y} = N_{v_z} = 32$\,cells \tabularnewline
    mass ratio & $\frac{m_i}{m_e} = 25$ \tabularnewline
    speed of light & $c = 20\,v_{\mathrm A}$ \tabularnewline
    \bottomrule
  \end{tabular}
\end{table}

In the simulation we use the $u_{z,s}$-based criterion with a thresholds
depicted in table \ref{tab:thresholds-uz}.
The criterion is reassessed every $0.1\,\omega_{c,i}^{-1}$ (inverse ion-gyrofrequencies)
for every subregion.
\begin{table}[h!]
  \caption{\label{tab:thresholds-uz} %
    Thresholds for the $u_z$-criterion in units of Alfv\'en-speed $v_{\mathrm
      A}$. They are evaluated for every subregion separately. The five-moment
    model is used iff neither the kinetic nor the ten-moment thresholds are met}
  \renewcommand{\arraystretch}{1.2}
  \begin{tabular}{l l l l}
    \toprule
    & & kinetic iff \dots & ten-moment iff not kinetic and \dots \tabularnewline
    \midrule \multirow{2}{*}{thresholds}
    & electrons & $\max \abs{u_{z,e}} \geq 0.3 v_{\mathrm A}$ & $\max \abs{u_{z,e}} \geq 0.1\,v_{\mathrm A}$ \tabularnewline
    & ions & $\max \abs{u_{z,i}} \geq 0.6 v_{\mathrm A}$ & $\max \abs{u_{z,i}} \geq 0.2\,v_{\mathrm A}$ \tabularnewline
    \bottomrule
  \end{tabular}
\end{table}

\begin{figure}[h!]
  \includegraphics{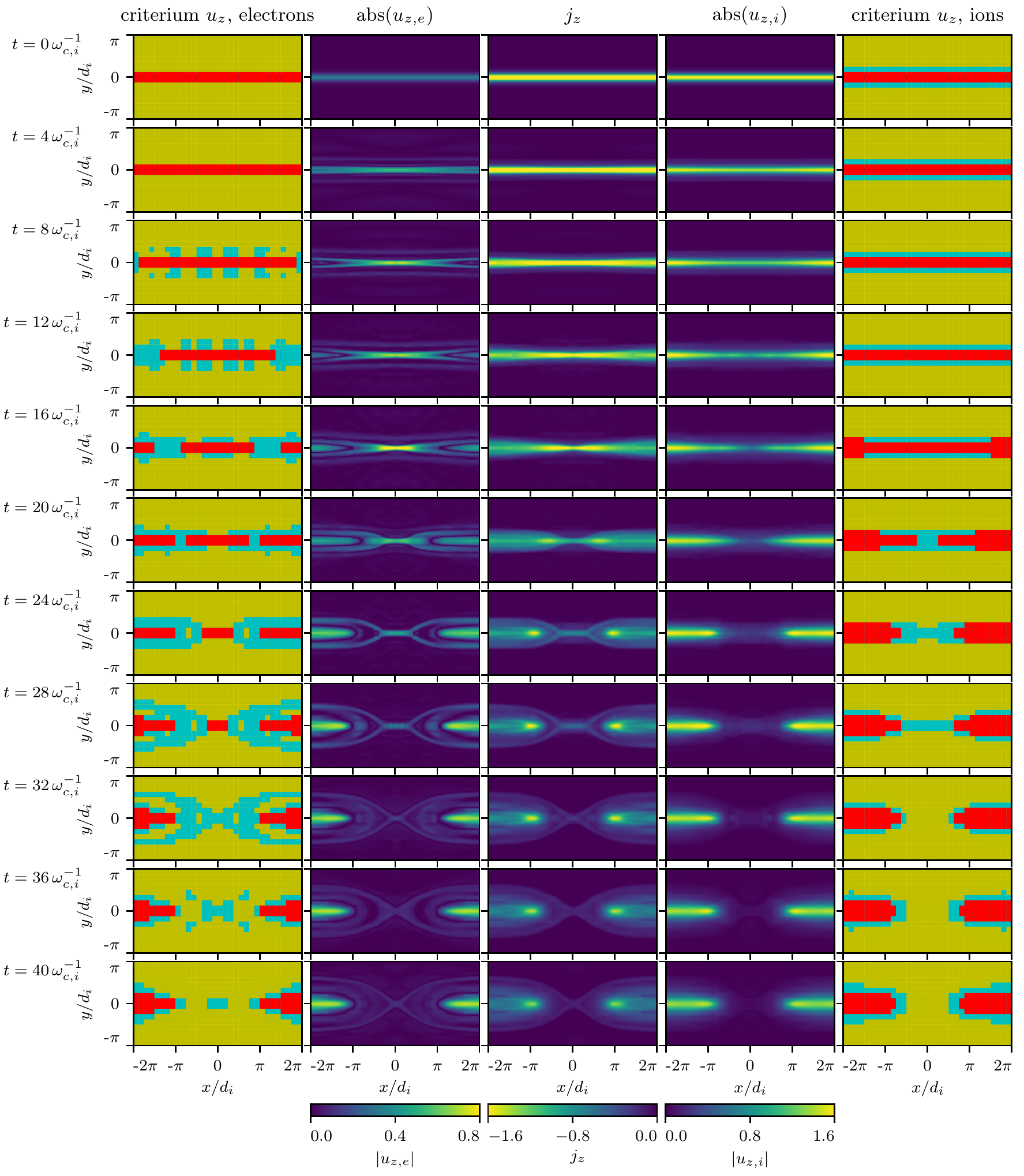}
  \caption{ \label{fig:time-series-uz1} %
    $\abs{u_{z,e}}$, $j_z$ and $\abs{u_{z,i}}$ for different times in units of
    inverse ion-gyrofrequencies $\omega_{c,i}^{-1}$. On the outsides, the used
    models are depicted based on the values of $u_{z,s}$ and threshold as given in
    table \ref{tab:thresholds-uz}. Red areas are solved with the kinetic solver,
    blue areas with the ten-moment and yellow areas with the five-moment fluid
    solver. Note that the initial models at time $t=0$ have been prescribed.}
\end{figure}

In figure \ref{fig:time-series-uz1} the fields $j_z$, $u_{z,e}$ and $u_{z,i}$
are shown together with the areas depicting the different physical models for
different times of the simulation. From these figures one can deduce that
substantial saving in simulation time can be achieved since the performance gain
is approximately proportional to the ratio of the computational domain to the
area where the Vlasov equation is solved.

\begin{figure}[h!]
  \includegraphics{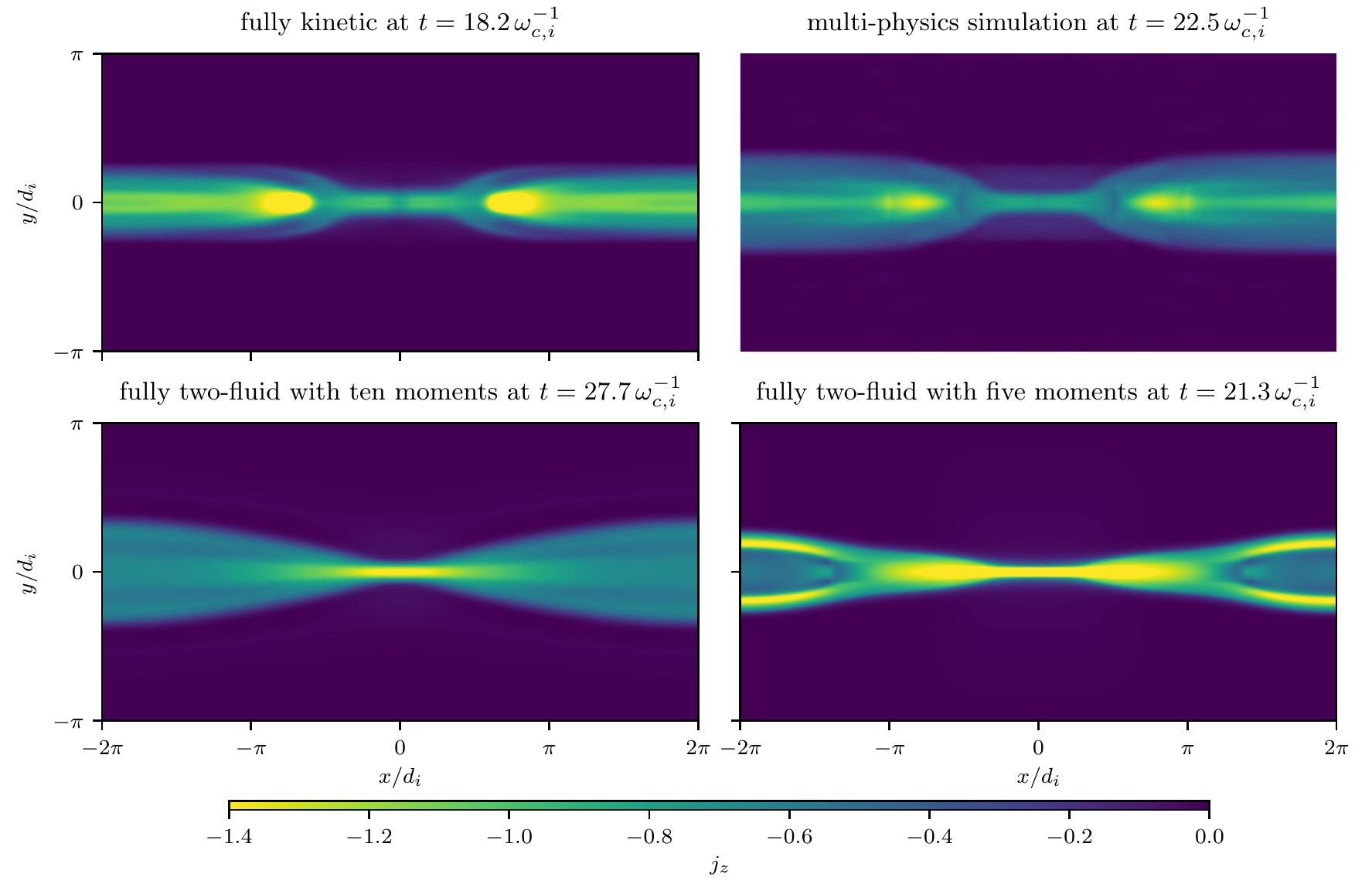}
  \caption{ \label{fig:model-comparison} %
    $j_z$ for the coupled run and some uncoupled comparison runs with a single
    solver}
\end{figure}

In figure \ref{fig:model-comparison}, a comparison of different simulations are
shown and compared to the multiphysics run. Depicted is the current density
$j_z$ at the time of the highest reconnection rate. The fully kinetic Vlasov
simulation agrees rather well with the multiphysics simulation. The overall
agreement is substantially better than the results obtained from the 10- and
5-moment simulations. However, differences especially in the precise values of
the absolute maxima of the current density, are visible. Whether this is an
effect of the model selection criteria has to be tested in further
investigations.

\begin{figure}[h!]
  \includegraphics{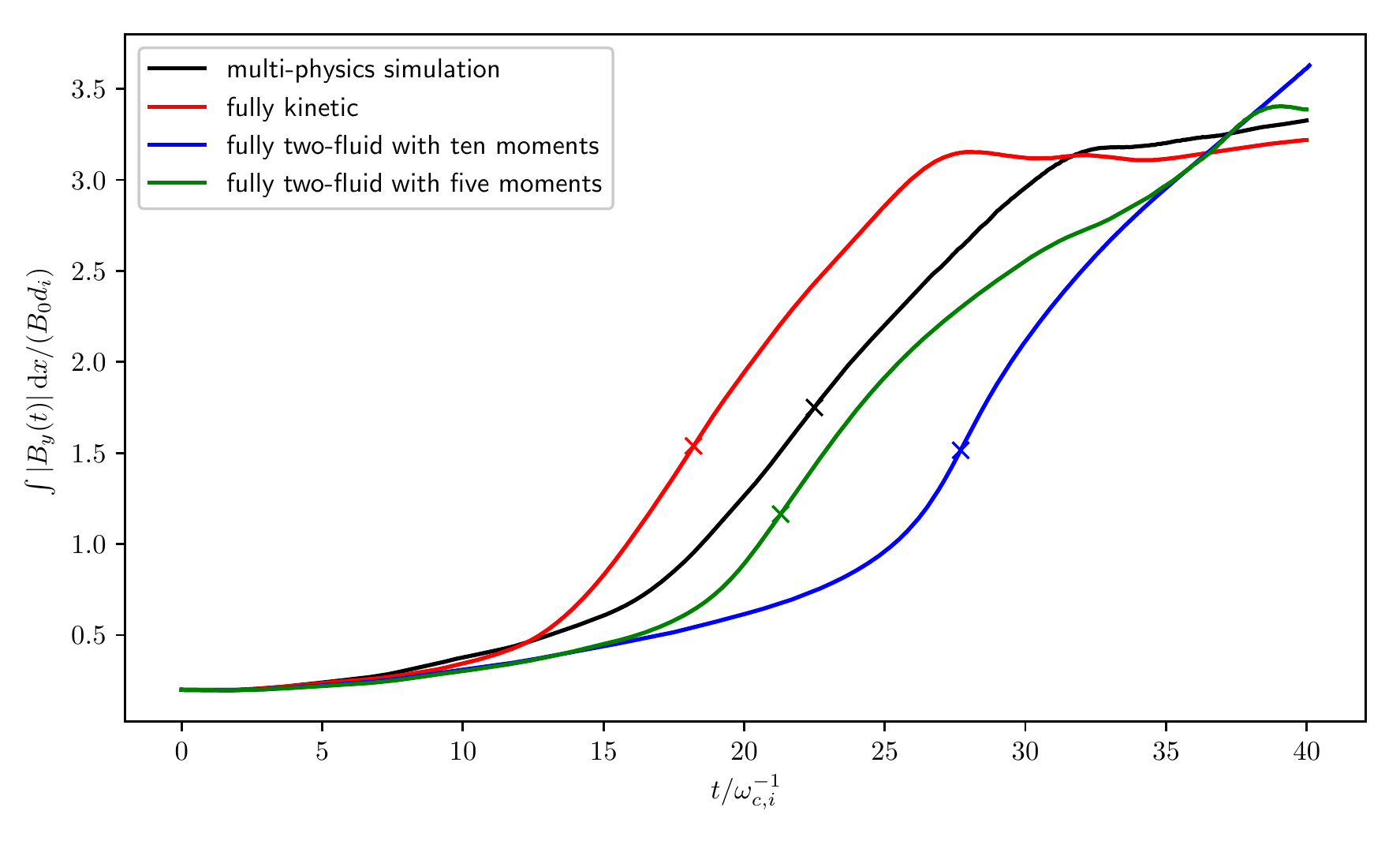}
  \caption{ \label{fig:reconnection-flux} %
    Reconnection fluxes for the coupled run and some uncoupled comparison runs
    with a single solver. Crosses mark the point of highest reconnection rate
    throughout the respective run}
\end{figure}

As a more quantitative comparison, the reconnecting flux for the multiphysics
simulation is plotted in figure \ref{fig:reconnection-flux} together with purely
kinetic and fluid runs. There are a number of things to observe from the plot:
While the reconnecting flux of the fluid runs does not saturate within the
simulation time, the kinetic and the multiphysics runs saturate. In addition,
they both saturate at the same level and thus capture essentially the same
small scale physics which is not possible with the fluid models.

Alone the presence of electron and ion kinetic regions in the very center of
reconnection zone and 10-moment fluid regions around it seems to ensure the
characteristic behaviour of the full kinetic reconnection scenario.

\section{Discussion}
We showed that the proposed multiphysics coupling hierarchy can give excellent
results even when only a small part of computational domain near the
reconnection zone is captured with a kinetic model.

However, still many questions and challenges remain and it is clear that the
present simulations are only on the level of a proof of concept.
Most important is the issue of designing robust physics refinement criteria and
their thresholds. First attempts based on the heat flux are under investigation.
In addition, the multiphysics coupling strategy should be formulated as an
asymptotic preserving scheme \cite{degond-deluzet:2017,hu-jin-li:2017}. The
coupling of the 10- and 5-moment models is already in this state when
incorporating the effect of Landau damping
\cite{wang-et-al:2015,allmann-rahn-trost-grauer:2018}. Presently, we are also
reformulating the coupling between the Vlasov and the 10-moment model. For this,
we formulate the kinetic description as an adaptive (in time and space) $\delta
f$ method and ease the transition to the fluid description as an asymptotic
preserving scheme. Finally, the multiphysics hierarchy should not stop at the
level of the 5-moment fluid description. Work to couple the 5-moment model to
MHD is in progress.

\section*{Acknowledgments}
We acknowledge interesting discussions with F. Allmann-Rahn, J. Dreher and
T. Trost.
Computations were conducted on the Davinci cluster at TP1 Plasma Research
Department and on the JURECA cluster at J\"ulich Supercomputing Center under the
project number HBO43.
The authors gratefully acknowledge the computing time granted by the John von
Neumann Institute for Computing (NIC) and provided on the supercomputer JURECA
at J\"ulich Supercomputing Centre (JSC).
% see http://www.fz-juelich.de/ias/jsc/EN/Expertise/Supercomputers/ComputingTime/Acknowledgements.html?nn=934984#doc1387592bodyText6

\bibliographystyle{frontiersinHLTH&FPHY} % for Health, Physics and Mathematics articles
\bibliography{lit}

\begin{thebibliography}{36}
\expandafter\ifx\csname natexlab\endcsname\relax\def\natexlab#1{#1}\fi
\expandafter\ifx\csname urlstyle\endcsname\relax
  \expandafter\ifx\csname doi\endcsname\relax
  \def\doi#1{doi:\discretionary{}{}{}#1}\fi \else
  \expandafter\ifx\csname doi\endcsname\relax
  \def\doi{doi:\discretionary{}{}{}\begingroup \urlstyle{rm}\Url}\fi \fi
\expandafter\ifx\csname selectlanguage\endcsname\relax
  \def\selectlanguage#1{}\fi

\bibitem[{Wang et~al.(2015)Wang, Hakim, Bhattacharjee, and
  Germaschewski}]{wang-et-al:2015}
Wang L, Hakim AH, Bhattacharjee A, Germaschewski K.
\newblock {Comparison of multi-fluid moment models with particle-in-cell
  simulations of collisionless magnetic reconnection}.
\newblock {\em Physics of Plasmas\/} {\bf 22} (2015) 012108.
\newblock \doi{http://dx.doi.org/10.1063/1.4906063}.

\bibitem[{Ng et~al.(2017)Ng, Hakim, Bhattacharjee, Stanier, and
  Daughton}]{ng-hakim-etal:2017}
Ng J, Hakim A, Bhattacharjee A, Stanier A, Daughton W.
\newblock Simulations of anti-parallel reconnection using a nonlocal heat flux
  closure.
\newblock {\em Physics of Plasmas\/} {\bf 24} (2017) 082112.
\newblock \doi{10.1063/1.4993195}.

\bibitem[{Allmann-Rahn et~al.(2018)Allmann-Rahn, Trost, and
  Grauer}]{allmann-rahn-trost-grauer:2018}
Allmann-Rahn F, Trost T, Grauer R.
\newblock Temperature gradient driven heat flux closure in fluid simulations of
  collisionless reconnection.
\newblock {\em accepted and to appear in J. Plasma Phys.\/}  (2018).

\bibitem[{Hammett and Perkins(1990)}]{hammett-perkins:1990}
Hammett GW, Perkins FW.
\newblock Fluid moment models for {L}andau damping with application to the
  ion-temperature-gradient instability.
\newblock {\em Phys. Rev. Lett.\/} {\bf 64} (1990) 3019--3022.
\newblock \doi{10.1103/PhysRevLett.64.3019}.

\bibitem[{Hammett et~al.(1992)Hammett, Dorland, and
  Perkins}]{hammett-dorland-perkins:1992}
Hammett GW, Dorland W, Perkins FW.
\newblock Fluid models of phase mixing, {L}andau damping, and nonlinear
  gyrokinetic dynamics.
\newblock {\em Physics of Fluids B\/} {\bf 4} (1992) 2052--2061.
\newblock \doi{http://dx.doi.org/10.1063/1.860014}.

\bibitem[{Passot and Sulem(2003)}]{passot-sulem:2003}
Passot T, Sulem PL.
\newblock Long-{A}lfv\'en-wave trains in collisionless plasmas. ii. a
  {L}andau-fluid approach.
\newblock {\em Physics of Plasmas\/} {\bf 10} (2003) 3906--3913.
\newblock \doi{10.1063/1.1600442}.

\bibitem[{Sharma et~al.(2003)Sharma, Hammett, and
  Quataert}]{sharma-hammett-etal:2003}
Sharma P, Hammett GW, Quataert E.
\newblock Transition from collisionless to collisional magnetorotational
  instability.
\newblock {\em The Astrophysical Journal\/} {\bf 596} (2003) 1121.

\bibitem[{Chust and Belmont(2006)}]{chust-belmont:2006}
Chust T, Belmont G.
\newblock Closure of fluid equations in collisionless magnetoplasmas.
\newblock {\em Physics of Plasmas\/} {\bf 13} (2006) 012506.
\newblock \doi{http://dx.doi.org/10.1063/1.2138568}.

\bibitem[{Schulze et~al.(2003)Schulze, Smereka, and E}]{Schulze2003}
Schulze TP, Smereka P, E W.
\newblock {Coupling kinetic Monte-Carlo and continuum models with application
  to epitaxial growth}.
\newblock {\em Journal of Computational Physics\/} {\bf 189} (2003) 197 -- 211.
\newblock \doi{http://dx.doi.org/10.1016/S0021-9991(03)00208-0}.

\bibitem[{Degond et~al.(2010)Degond, Dimarce, and Mieussens}]{deg2010}
Degond P, Dimarce G, Mieussens L.
\newblock {A multiscale kinetic-fluid solver with dynamic localization of
  kinetic effects}.
\newblock {\em J. Comput. Phys.\/} {\bf 229} (2010) 4907--4933.

\bibitem[{Dellacherie(2003)}]{del2003}
Dellacherie S.
\newblock Kinetic-fluid coupling in the field of the atomic vapor isotopic
  separation: Numerical results in the case of a monospecies perfect gas.
\newblock {\em AIP Conf. Proc.\/} {\bf 663} (2003) 947--956.
\newblock \doi{10.1063/1.1581642}.

\bibitem[{Goudon et~al.(2013)Goudon, Jin, Liu, and Yan}]{gou2013}
Goudon T, Jin S, Liu JG, Yan B.
\newblock Asymptotic-preserving schemes for kinetic-fluid modeling of disperse
  two-phase flows.
\newblock {\em J. Comput. Phys.\/} {\bf 246} (2013) 145--164.

\bibitem[{Tiwari and Klar(1998)}]{tiwari-klar:1998}
Tiwari S, Klar A.
\newblock {An adaptive domain decomposition procedure for Boltzmann and Euler
  equations}.
\newblock {\em J. Comput. Appl. Math.\/} {\bf 90} (1998) 223--237.
\newblock \doi{http://dx.doi.org/10.1016/S0377-0427(98)00027-2}.

\bibitem[{Le~Tallec and Mallinger(1997)}]{Tal1997}
Le~Tallec P, Mallinger F.
\newblock {Coupling Boltzmann and Navier-Stokes Equations by Half Fluxes}.
\newblock {\em Journal of Computational Physics\/} {\bf 136} (1997) 51–67.
\newblock \doi{10.1006/jcph.1997.5729}.

\bibitem[{Sugiyama and Kusano(2007)}]{Sug2007}
Sugiyama T, Kusano K.
\newblock Multi-scale plasma simulation by the interlocking of
  magnetohydrodynamic model and particle-in-cell kinetic model.
\newblock {\em Journal of Computational Physics\/} {\bf 227} (2007)
  1340–1352.
\newblock \doi{10.1016/j.jcp.2007.09.011}.

\bibitem[{Markidis et~al.(2014)Markidis, Henri, Lapenta, R\"onnmark, Hamrin,
  Meliani et~al.}]{Mar2014}
Markidis S, Henri P, Lapenta G, R\"onnmark K, Hamrin M, Meliani Z, et~al.
\newblock The fluid-kinetic particle-in-cell method for plasma simulations.
\newblock {\em Journal of Computational Physics\/}  (2014).
\newblock \doi{10.1016/j.jcp.2014.02.002}.

\bibitem[{Daldorff et~al.(2014)Daldorff, T{\'o}th, Gombosi, Lapenta, Amaya,
  Markidis et~al.}]{daldorff-et-al:2014}
Daldorff LKS, T{\'o}th G, Gombosi TI, Lapenta G, Amaya J, Markidis S, et~al.
\newblock {Two-way coupling of a global Hall magnetohydrodynamics model with a
  local implicit particle-in-cell model}.
\newblock {\em J. Comput. Phys.\/} {\bf 268} (2014) 236--254.
\newblock \doi{10.1016/j.jcp.2014.03.009}.

\bibitem[{Kolobov and Arslanbekov(2012)}]{Kol2012}
Kolobov V, Arslanbekov R.
\newblock Towards adaptive kinetic-fluid simulations of weakly ionized plasmas.
\newblock {\em Journal of Computational Physics\/} {\bf 231} (2012) 839–869.
\newblock \doi{10.1016/j.jcp.2011.05.036}.

\bibitem[{Rieke et~al.(2015)Rieke, Trost, and Grauer}]{rieke-et-al:2015}
Rieke M, Trost T, Grauer R.
\newblock {Coupled Vlasov and two-fluid codes on GPUs}.
\newblock {\em J. Comput. Phys.\/} {\bf 283} (2015) 436--452.
\newblock \doi{10.1016/j.jcp.2014.12.016}.

\bibitem[{Birn et~al.(2001)Birn, Drake, and Shay}]{birn2001}
Birn J, Drake JF, Shay MA.
\newblock {Geospace Environmental Modeling (GEM) Magnetic Reconnection
  Challenge}.
\newblock {\em J. Geophys. Res.\/} {\bf 106} (2001) 3715--3719.

\bibitem[{Strang(1968)}]{strang:1968}
Strang G.
\newblock {On the construction and comparison of difference schemes}.
\newblock {\em SIAM J. Numer. Anal.\/} {\bf 5} (1968) 506--517.

\bibitem[{Filbet et~al.(2001)Filbet, Sonnendr{\"u}cker, and
  Bertrand}]{filbet2001}
Filbet F, Sonnendr{\"u}cker E, Bertrand P.
\newblock Conservative numerical schemes for the {V}lasov equation.
\newblock {\em J. Comput. Phys.\/} {\bf 172} (2001) 166--187.

\bibitem[{Leslie and Purser(1995)}]{leslie-purser:1995}
Leslie LM, Purser RJ.
\newblock {Three-dimensional mass-conserving semi-Lagrangian scheme employing
  forward trajectories}.
\newblock {\em Monthly Weather Review\/} {\bf 123} (1995) 2551--2566.

\bibitem[{Boris(1970)}]{boris:1970}
Boris JP.
\newblock Relativistic plasma simulation-optimization of a hybrid code.
\newblock {\em Proc. Fourth Conf. Num. Sim. Plasmas, Naval Res. Lab, Wash.
  DC\/} (1970), 3--67.

\bibitem[{Schmitz and Grauer(2006{\natexlab{a}})}]{schmitz-grauer:2006a}
Schmitz H, Grauer R.
\newblock {Darwin–Vlasov simulations of magnetised plasmas}.
\newblock {\em J. Comput. Phys.\/} {\bf 214} (2006{\natexlab{a}}) 738--756.
\newblock \doi{10.1016/j.jcp.2005.10.013}.

\bibitem[{Schmitz and Grauer(2006{\natexlab{b}})}]{schmitz-grauer:2006b}
Schmitz H, Grauer R.
\newblock {Comparison of time splitting and backsubstitution methods for
  integrating {V}lasov's equation with magnetic fields}.
\newblock {\em Comp. Phys. Commun.\/} {\bf 175} (2006{\natexlab{b}}) 86--92.
\newblock \doi{10.1016/j.cpc.2006.02.007}.

\bibitem[{Forum(1994)}]{MPI:1994}
Forum MP.
\newblock {MPI}: A message-passing interface standard.
\newblock Tech. rep., Knoxville, TN, USA (1994).

\bibitem[{n{V}idia corp.(2013)}]{programming-guide}
[Dataset] n{V}idia corp.
\newblock n{V}idia {CUDA} {C} programming guide.
\newblock \url{http://docs.nvidia.com/cuda/cuda-c-programming-guide/index.html}
  (2013).

\bibitem[{Kurganov and Levy(2000)}]{kurganov-levy:2000}
Kurganov A, Levy D.
\newblock {A Third-Order Semidiscrete Central Scheme for Conservation Laws and
  Convection-Diffusion Equations}.
\newblock {\em SIAM J. Sci. Comput.\/} {\bf 22} (2000) 1461--1488.

\bibitem[{Shu and Osher(1988)}]{shu-osher:1988}
Shu CW, Osher S.
\newblock {Efficient Implementation of Essentially Non-oscillatory
  Shock-Capturing Schemes}.
\newblock {\em J. Comput. Phys.\/} {\bf 77} (1988) 439--471.

\bibitem[{Yee(1966)}]{yee:1966}
Yee KS.
\newblock {Numerical Solution of Initial Boundary Value Problems Involving
  Maxwell's Equations in Isotropic Media}.
\newblock {\em IEEE Trans. Antennas Propag.\/} {\bf 14} (1966) 302--307.

\bibitem[{Taflove(1995)}]{taflove1995}
Taflove A.
\newblock {\em Computational Electrodynamics: The Finite-Difference Time-Domain
  Method\/} (Artech House) (1995).

\bibitem[{Dreher and Grauer(2005)}]{dreher-grauer:2005}
Dreher J, Grauer R.
\newblock \textit{racoon}: A parallel mesh-adaptive framework for hyperbolic
  conservation laws.
\newblock {\em Parallel Computing\/} {\bf 31} (2005) 913.

\bibitem[{{J\"{u}lich Supercomputing Centre}(2016)}]{jureca}
{J\"{u}lich Supercomputing Centre}.
\newblock {JURECA: General-purpose supercomputer at J\"{u}lich Supercomputing
  Centre}.
\newblock {\em Journal of large-scale research facilities\/} {\bf 2} (2016).
\newblock \doi{10.17815/jlsrf-2-121}.

\bibitem[{Degond and Deluzet(2017)}]{degond-deluzet:2017}
Degond P, Deluzet F.
\newblock Asymptotic-preserving methods and multiscale models for plasma
  physics.
\newblock {\em Journal of Computational Physics\/} {\bf 336} (2017) 429 -- 457.
\newblock \doi{https://doi.org/10.1016/j.jcp.2017.02.009}.

\bibitem[{Hu et~al.(2017)Hu, Jin, and Li}]{hu-jin-li:2017}
Hu J, Jin S, Li Q.
\newblock Chapter 5 - asymptotic-preserving schemes for multiscale hyperbolic
  and kinetic equations.
\newblock Abgrall R, Shu CW, editors, {\em Handbook of Numerical Methods for
  Hyperbolic Problems\/} (Elsevier), {\em Handbook of Numerical Analysis\/},
  vol.~18 (2017), 103 -- 129.
\newblock \doi{https://doi.org/10.1016/bs.hna.2016.09.001}.

\end{thebibliography}

\end{document}